\documentclass[12pt,a4paper]{article} 

\usepackage{epsfig}
\usepackage{amssymb}

\newcommand{\be}{\begin{equation}} 
\newcommand{\en}{\end{equation}}
\newcommand{\bea}{\begin{eqnarray}}
\newcommand{\ena}{\end{eqnarray}}

\newcommand{\hbo}{\hbox to 1 true cm {\hfill } } 
\newcommand{\tr}{\hbox{tr}}

\def\dslash{\partial\kern-.6em\slash}
\def\kslash{k\kern-.5em\slash}
\def\pslash{p\kern-.4em\slash}
\def\Dslash{D\kern-.6em\slash}
\def\Vslash{V\kern-.7em\slash}
\def\vslash{v\kern-.5em\slash}
\def\rslash{r\kern-.5em\slash}
\def\qslash{q\kern-.5em\slash}

\begin{document} 
\vglue 1truecm
  
\vbox{ UNITU-THEP-7/2000  
\hfill July 26, 2000
}
  
\vfil
\centerline{\large\bf Center-vortex dominance after dimensional reduction } 
\centerline{\large\bf of $SU(2)$ lattice gauge theory } 
  
\bigskip
\centerline{ J.~Gattnar, K.~Langfeld, A.~Sch\"afke$^a$, H.~Reinhardt$^b$ } 
\vspace{1 true cm} 
\centerline{ Institut f\"ur Theoretische Physik, Universit\"at 
   T\"ubingen }
\centerline{D--72076 T\"ubingen, Germany}
  
\vfil
\begin{abstract}
The high-temperature phase of $SU(2)$ Yang-Mills theory is addressed 
by means of dimensional reduction with a special emphasis on 
the properties of center vortices. For this purpose, the vortex vacuum 
which arises from center projection is studied in pure 
3-dimensional Yang-Mills theory as well as in the 3-dimensional 
adjoint Higgs model which describes the high temperature phase 
of the 4-dimensional $SU(2)$ gauge theory.  We find center-dominance 
within the numerical accuracy of 10\%.

\end{abstract}

\vfil
\hrule width 5truecm
\vskip .2truecm
\begin{quote} 
$^a$ supported in part by Graduiertenkolleg {\it Hadronen und Kerne} 

$^b$ supported in part by grant DFG Re 856/4-1

\end{quote}
\eject

\centerline{\bf 1. Introduction: \hfill } 

The idea that vortices are the degrees of freedom which are responsible 
for confinement dates back to the pioneering work of 't~Hooft, 
Aharonov et al.~\cite{tho78} and Mack et al.~\cite{mack}. 
By introducing twisted boundary conditions, 't~Hooft introduced 
topologically stable center vortices winding around the torus 
(space-time manifold) in order to test the response of the 
Yang-Mills system to the imprinted magnetic flux. 
The free energy of such a flux may serve as an order parameter 
for (de-)confinement and was explicitly calculated in a recent lattice 
study~\cite{kov00} which is based on the method of reference~\cite{hoe99}. 
Mack and collaborators explicitly extracted 
center degrees of freedom from the plaquettes and studied the 
dynamics of the emerging center vortices~\cite{mack}. In particular, 
it was observed that randomly distributed center vortices give 
rise to an area law for the Wilson loop, hence, implying confinement. 
The same definition of center degrees of freedom was subsequently 
resumed in~\cite{kov98} and investigated by lattice calculations. 

\vskip 0.3cm 
The vortex picture of confinement~\cite{tho78,mack} has recently 
experienced a revival due to the observation of center dominance 
of the string tension in the so-called maximal center gauge~\cite{deb98}. 
After bringing lattice configurations into this gauge, one projects 
each link to its nearest center element (center projection), thus 
obtaining an effective $Z_2$ gauge, i.e. the center 
vortex\footnote{ this vortex type must not be confused with the 
vortex types discussed in~\cite{tho78} and in~\cite{mack}.} theory, which 
accounts for the full string tension~\cite{deb98,la98}. 
Furthermore, their (area) 
density as well as their interactions turned out to be meaningful 
quantities in the continuum limit~\cite{la98} (see also~\cite{deb98}). 
In addition, the latter 
vortex picture also provides an appealing picture of the deconfinement 
phase transition at finite temperatures: the vortex ensemble undergoes 
a de-percolation transition from a phase of percolating vortices at 
low temperatures to a phase of small vortex clusters at high 
temperatures~\cite{la99}. In fact, the de-percolation transition 
is seen in the 3-dimensional hypercubes of the 4-dimensional lattice 
universe, arising at fixed space slices. 
In these hypercubes, the vortices partially align parallel to 
the time axis~\cite{la99}. On the other hand, 
the vortices which are detected in the spatial hypercube at a given time 
are still percolating even at high temperatures. 

\vskip 0.3cm 
A detailed understanding of the high temperature phase of Yang-Mills 
theory is highly desirable for understanding signatures of the 
quark gluon plasma as it might be produced in near future collider 
experiments at RHIC and LHC. 
It was proposed in the early eighties~\cite{app81} that at asymptotic 
temperatures $T$, 4-dimensional Yang-Mills theory effectively 
reduces to the 3-dimensional counterpart coupled to adjoint 
Higgs matter~\cite{lac92,kar94}. The (dimensionful) 
coupling constant of the latter theory becomes 
\be 
\frac{1}{g_3^2 } \; = \; \frac{1}{ T \, g^2(T)} . 
\label{eq:1} 
\en 
Thereby $g^2(T)$ denotes the 4-dimensional running coupling constant. 
Since the dimensional reduction of 4-dimensional Yang-Mills theory yields 
the 3-dimensional adjoint Higgs Yang-Mills theory in the 
confining phase~\cite{kar94}, one expects that the spatial string tension 
$\sigma _s $ (of 4-dimensional) Yang-Mills theory scales with the 
dimensionful parameter $g_3^4 = g^4(T) \; T^2 $. Indeed, a large scale 
numerical analysis~\cite{bali93} yields 
\be 
\sigma _s (T) \; = \; c \, g_3^4 \; = \; c \, g^4(T) \; T^2 \; , 
\hbo c \; = \; 0.136 \pm 0.011 \; . 
\label{eq:2} 
\en

\vskip 0.3cm 
In this letter, we investigate the high temperature phase of $SU(2)$ 
Yang-Mills theory in the center vortex picture defined in~\cite{deb98}. 
In a previous paper~\cite{la99}, it was shown that the 
spatial string tension which is calculated after vortex projection 
increases with increasing temperature according the expectations 
of dimensional reduction (\ref{eq:2}). In order to show that 
the center vortex scenario is a sensible description of the Yang-Mills 
vacuum at high temperatures and to reveal the mechanism of 
dimensional reduction in the vortex picture it is essential to demonstrate 
that the center vortices of the {\it 3-dimensional } Yang-Mills theory 
coupled with adjoint Higgs matter survive the continuum limit 
and that a projection of the full 3-dimensional theory onto the 
center vortex vacuum accounts for the full (spatial) string tension. 
We will find an agreement of projected and un-projected 
string tension within the achieved numerical accuracy of 10\%.

\vskip 0.3cm 
\centerline{\bf 2. Center projection in the high temperature limit: \hfill } 

For extracting the structure of the two dimensional vortex 
world sheets in four space-time dimensions, we adopt the 
maximal center gauge 
and subsequently perform center projection~\cite{deb98}. 
Let $U_\mu (x)$, $\mu =0 \ldots 3$ denote 
the $SU(2)$ link variables and $\Omega (x)$ a gauge transformation. 
The maximal center gauge condition amounts to maximizing the functional 
\be 
S_{\mathrm fix} \; = \; \sum _{\{x\},\mu } \biggl[ \tr U^\Omega _\mu (x) \biggr]^2 
\rightarrow \hbox{max} \; , \hbo 
U^\Omega _\mu  (x) = \Omega (x) U_\mu (x) \Omega ^\dagger (x+\mu). 
\label{eq:4} 
\en 
with respect to $\Omega (x)$. For calculating the gauge matrices $\Omega (x)$ 
for a given link configuration $U_\mu (x) $ we use an iteration 
over-relaxation algorithm~\cite{deb98}. Although there is no conceptual 
Gribov problem with the gauge condition (\ref{eq:4}), one encounters 
a {\it practical} Gribov problem~\cite{kov99,born00} when looking for the 
global maximum of $S_{\mathrm fix}$ with numerical techniques. The practical 
Gribov problem can be alleviated by changing the gauge condition 
(\ref{eq:4}) to its Laplacian version~\cite{vin92,ale99} or 
to alternative modifications of the maximal center gauge as e.g.~proposed 
in~\cite{la00}. 
We believe, however, that the naive iteration over-relaxation algorithm 
is capable to grasp the essential physics of the 
center vortex vacuum and relegate an investigation of the center vortex 
properties in Laplacian gauge to future work. 
Once the gauge condition (\ref{eq:4}) is installed, center-projection 
of $SU(2) \rightarrow Z_2$ is performed by replacing the gauge 
fixed link variables by their closest center element 
\be 
U^\Omega _\mu  (x) \; \rightarrow \; \hbox{sign} \, \bigl\{ \tr 
U^\Omega _\mu (x) \bigr\} \; \in \; \{-1, +1 \} \; . 
\label{eq:5} 
\en 

\vskip 0.3cm 
In the following, we will establish a relation between the 
4-dimensional vortex world sheets of $SU(2)$ Yang-Mills theory at high 
temperatures and the vortex world lines of the dimensionally reduced 
3-dimensional theory. For this purpose, we first derive the gauge constraint 
for gauge transformations $\Omega (\vec{x})$ of the spatial hypercube. 
Decomposing (\ref{eq:4}) as 
\be 
S_{\mathrm fix} \; = \; \sum _{\{x\},k } \biggl[ \tr U^\Omega _k (x) \biggr]^2 
\; + \;  \sum _{\{x\} } \biggl[ \tr U^\Omega _0 (x) \biggr]^2 \; , \hbo 
k=1 \ldots 3 \; , 
\label{eq:4a} 
\en 
and using the fact that $\tr U_0(x)$ is invariant under time independent 
gauge transformations, the gauge condition (\ref{eq:4}) seamless 
extends to three dimensions, i.e. 
\be 
S^\prime_{\mathrm fix} 
\; = \; \sum _{\{x\},k } \biggl[ \tr U^\Omega _k (x) \biggr]^2 
\rightarrow \hbox{max} \; , \hbo k=1\ldots 3 \; ,
\hbo \Omega = \Omega (\vec{x}) \; . 
\label{eq:4b} 
\en 

\vskip 0.3cm 
Constructing the dimensionally reduced theory, the integration over 
the link variables $U_0(x)$ is replaced by an integration over the 
field $A_0(x)$ which lives in the algebra~\cite{lac92}, 
$ A_0 \in \mathfrak{su}(2) 
\widehat{=} SO(3) \widehat{=} SU(2) / Z_2 $, disregarding 
the center $Z_2$. 
In the vortex picture, this assumption is consistent with the observation 
that the vortex world sheets of the high temperature phase are 
not linked with time-like Wilson loops~\cite{la99}. 

\vskip 0.3cm 
The assumption which is inherent in the construction of the dimensionally 
reduced theory is that the vortex world sheets of high temperature Yang-Mills 
theory are aligned along the time axis and wrap 
around the torus in time direction. This assumption is supported by 
lattice calculations~\cite{la99}. The vortex world lines 
which are detected in a spatial hypercube at a given time slice 
can be obtained by the 3-dimensional version of the maximal center gauge 
fixing condition (\ref{eq:4b}) and by standard 
center projection which consider spatial link variables only.

\vskip 0.3cm 
\centerline{\bf 3. Three-dimensional pure $SU(2)$ gauge theory: \hfill } 

Resorting to pure 3-dimensional Yang-Mills theory (i.e.~without 
adjoint Higgs matter), one 
obtains in the continuum limit~\cite{tep99}
\be 
\sigma _s \; \approx \; 0.11 \; g_3^4 \; , \hbox to 6 cm 
{\hfill (pure 3-D YM-theory) } \; . 
\label{eq:2a} 
\en 
This value is remarkably close to the value (\ref{eq:2}) obtained in the full 
4-dimen\-sional theory. This indicates that the 
static quark correlations within the spatial bulk are dominated 
by the 3-dimensional Yang-Mills theory. Note, however, that 
the quark {\it potential } of the 4-dimensional Yang-Mills theory 
at high temperatures is related to correlations of the adjoint Higgs field 
and sensitively depends on the Higgs gauge field couplings~\cite{lac92}. 

\vskip 0.3cm 
In a first step, we will investigate 3-dimensional pure $SU(2)$ Yang-Mills 
theory with Wilson action and we will neglect the coupling to the adjoint 
Higgs matter. 
This will be an approximation of the dimensionally 
reduced 4-dimensional Yang-Mills theory, but will shed light onto 
the confinement mechanism of 3-dimensional $SU(2)$ gauge theory, which 
is an interesting issue on its own. 

\vskip 0.3cm 
In this section, we will calculate the static quark anti-quark 
potential in the projected and the un-projected theory as well 
as the center vortex area density and we will extrapolate the data to 
continuum limit. The (spatial) string tension $\sigma _s$ in units 
of the lattice spacing $a$ is a function of $\beta = 4/g_3^2 a $,  
the only parameter of the theory. High statistics Monte-Carlo 
simulations~\cite{tep99} show 
\be 
\sigma _s \, a^2 \; = \; \frac{1.788}{\beta ^2 } \left( 
1 \, + \, \frac{1.414}{\beta } \, + \, \ldots \right) 
\hbox to 2cm {\hfill for \hfill } \beta \ge 3 \; . 
\label{eq:3} 
\en 
The continuum limit is approached by taking the limit $\beta 
\rightarrow \infty $ at a fixed value of the reference scale $\sigma _s$. 

\begin{figure}[t]
\centerline{ 
\epsfxsize=10cm
\epsffile{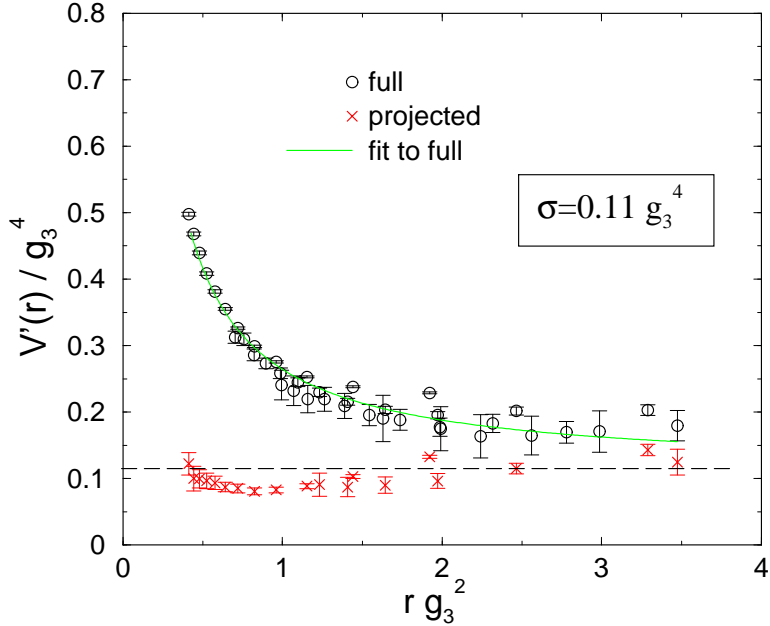}
}
\caption{ The derivative of the static quark anti-quark potential of the 
   full 3-dimensional theory and calculated from $Z_2$-projected links.} 
\label{fig:1} 
\end{figure}
\vskip 0.3cm

\begin{figure}[t]
\centerline{ 
\epsfxsize=10cm
\epsffile{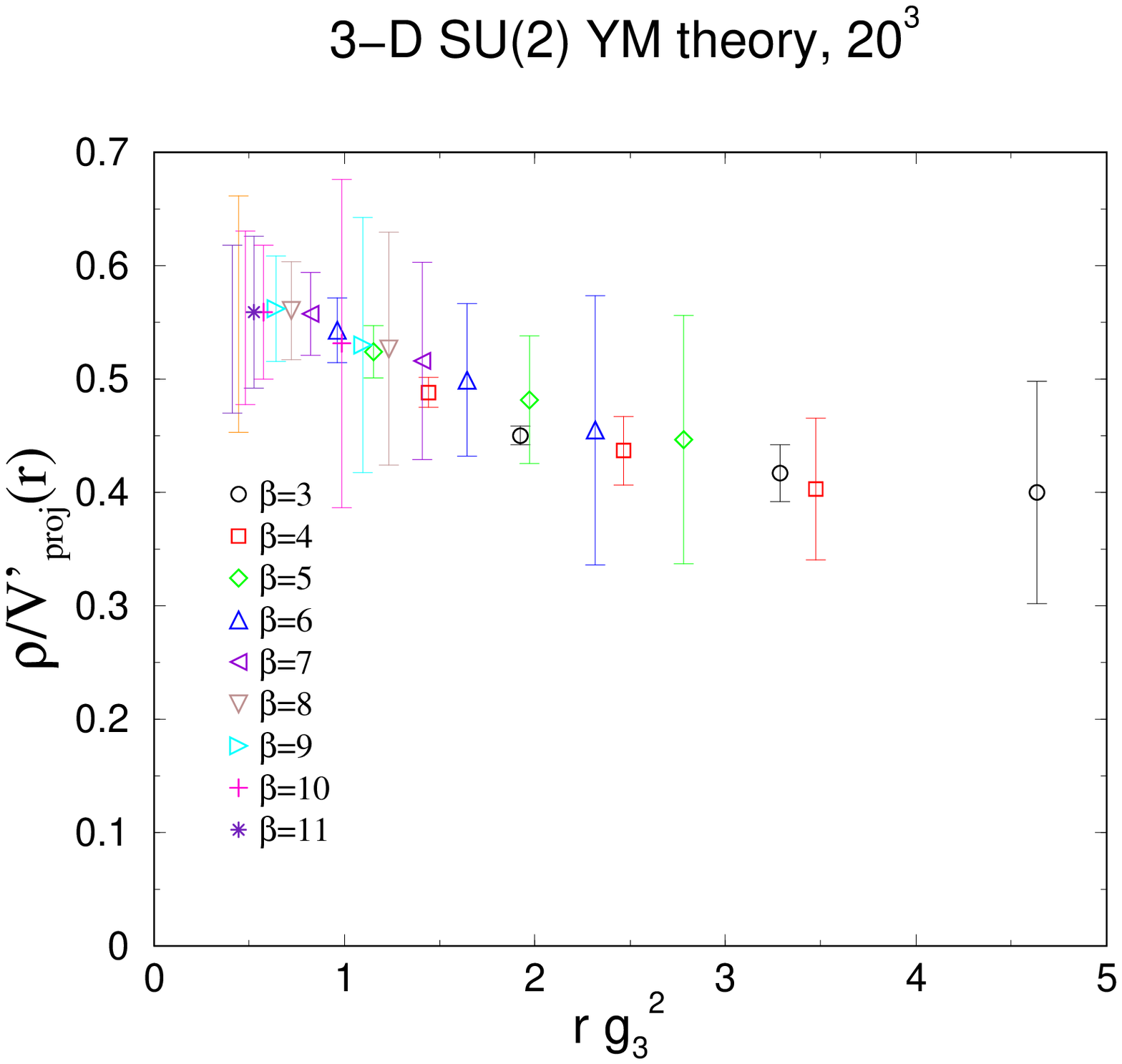}
}
\caption{ The ratio of the  vortex area density $\rho $ and the 
   derivative of the projected quark anti-quark potential at several 
   renormalization points $\beta $. } 
\label{fig:2} 
\end{figure}
\vskip 0.3cm 
In a first step, we compared the static quark anti-quark 
potential $V_{\mathrm proj} (r)$ as a function of the 
distance $r$ of the quark anti-quark pair with the 
result which is calculated from the projected link variables 
(for the derivative of this potential see figure~\ref{fig:1}). 
The calculation were carried out on $20^3$ lattice, and three attempts 
were made to find the global maximum of (\ref{eq:4b}) with the iteration 
over-relaxation algorithm. The fit $V^\prime (r) = \sigma _s + a/r $ 
to the full result yields $\sigma _s = 0.11 \, g_3^4 $ which 
is in perfect agreement with data presented in~\cite{tep99}. 
We observe a slight increase of the projected potential with 
increasing $r$. The (spatial) string tension obtained from the 
projected potential is in agreement with the full (spatial) string tension 
within statistical errors. 
We recover the same qualitative behavior of the potential as we did 
in the case of the 4-dimensional theory: the short distance behavior 
due to gluon radiation is changed by projection while the long range 
physics of the potential is roughly un-changed. This signals 
center vortex dominance of the (spatial) string tension in 3-dimensional 
Yang-Mills theory. 

\vskip 0.3cm 
In a second step, we rephrase the $Z_2$ gauge theory which is 
obtained by projection (\ref{eq:5}) in terms of vortices. 
A vortex is said to pierce an elementary plaquette if the 
plaquette calculated with $Z_2$ links yields $(-1)$. One then 
shows by virtue of the $Z_2$ Bianchi identity that this vortex 
material forms closed loops (in three dimension). Guided by the 
4-dimensional investigation~\cite{la98}, we investigate whether 
these vortices extrapolate to the continuum limit $a\rightarrow 0$ 
by calculating the vortex area density $\rho a^2$ for large values 
of $\beta $. Since it is believed that the vortex area density 
is the relevant scale for the projected quark anti-quark potential, 
we calculated the ratio of this density and the derivative of the 
projected potential, i.e. $V^\prime _{\mathrm proj} (r)$, 
as a function of $r$. 
This derivative might extrapolate to the value of the full string tension 
(see figure \ref{fig:1}). The result is shown in figure \ref{fig:2}. 
The data for this ratio are slightly decreasing for increasing $r$. 
The ratio, however, seems to approach a constant value for large values 
of $r$, hence, suggesting that the vortex area density extrapolates 
to the continuum limit. 

\vskip 0.3cm 
Let us compare the asymptotic value of that ratio with the corresponding 
value calculated in 
4-dimensional Yang-Mills theory at $T \approx 2 T_c$. Using $\rho _s$ 
reported in~\cite{la99} and the spatial string tension provided 
in~\cite{bali93}, one finds 
\be 
\rho / \sigma _s \; \approx \; 0.33 \; , 
\hbox to 8cm {\hfill (4-D Yang-Mills theory, $T \approx 2T_c$) } \; . 
\label{eq:5a} 
\en 
There is an agreement of the ratio (\ref{eq:5a}) with the corresponding 
value estimated from 3-dimensional Yang-Mills theory at the 
10\% level.

\begin{figure}[t]
\centerline{ 
\epsfxsize=10cm
\epsffile{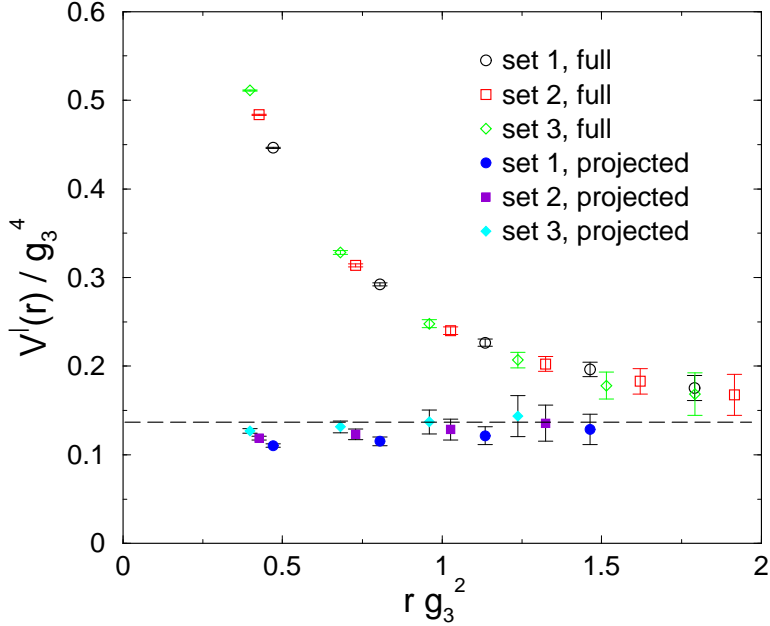}
}
\caption{ The derivative of the (spatial) static quark potential  
   calculated with the dimensionally reduced theory. The dashed line 
   indicates the spatial string tension (\ref{eq:2}) of the 
   full 4-dimensional theory.  } 
\label{fig:3} 
\end{figure}
\vskip 0.3cm 
\centerline{\bf 4. Three-dimensional $SU(2)$ adjoint Higgs theory: \hfill } 

\begin{table}[t] 
\caption{ Parameter sets of the effective dimensionally reduced theory for 
   a $24^3$ lattice taken from \cite{lac92}. } 
\label{tab:1}
$$ \hbox{ 
\begin{tabular}{cccccc} \hline
 & $T/T_c$  &  $ \beta _4 $ & $\beta $ & $h$     & $\kappa $ \\ \hline 
set 1 & $2.0$    &  $2.50$       & $12.25$  & $-.30 $ & $0.106 $ \\ 
set 2 & $3.5$    &  $2.80$       & $13.54$  & $-.26 $ & $0.094 $ \\ 
set 3 & $6.0$    &  $3.00$       & $14.48$  & $-.24 $ & $0.086 $ \\ \hline 
\end{tabular} 
} $$
\end{table} 
It was observed in~\cite{lac92} that static quark anti-quark potential 
of the high temperature phase of 4-dimensional is well reproduced 
be an effective 3-dimensional theory described in terms of the 
action
\be 
S_{\mathrm eff} \; = \; S_{\mathrm YM}(U) \, + \, S_{\mathrm hop} 
(U,A_0) \, + \, S_{\mathrm sc} (A_0) \; , 
\label{eq:10} 
\en 
where $S_{YM}$ is the 3-dimensional Wilson action. The field $A_0$ 
lives in the space of the $SU(2)$ algebra. Its is a reminder of the 
zeroth component of the 4-dimensional gauge fields $A_\mu $, 
$\mu = 0 \ldots 3$. It 
couples like an adjoint Higgs field in the reduced theory~\cite{lac92}, 
i.e. 
\bea 
S_{\mathrm hop} (U,A_0) &=& - \frac{1}{2} \beta \sum_{\{x\}, k } 
\tr \biggl[ A_0(x) \, U_k (x) \, A_0(x+k) \, U^\dagger _k (x) 
\biggr] \; , 
\label{eq:11} \\ 
S_{\mathrm sc} (A_0) &=& \frac{1}{2} \beta \sum_{\{x\}} \left\{ 
(3 + h/2) \, \tr A_0^2 \; + \; \kappa \left[ \frac{1}{2} \tr A_0^2(x) 
\right]^2 \right\} \; . 
\label{eq:12} 
\ena
We use hermitian Pauli matrices as generators of the $SU(2)$ algebra. 
The dimensionally reduced theory is thus described by the partition 
function 
\be 
Z_{\mathrm eff} \; = \; \int {\cal D} U \; {\cal D} A_0 \; \exp \biggl\{ 
-S_{\mathrm eff}(U,A_0) \biggr\} \; . 
\label{eq:13} 
\en
The integration over the link variables $U$ (of the 3-dimensional lattice) 
takes into account the Haar measure, while the integration over 
the $A_0$ field is carried out with a flat measure. 
A comparison of the static potential calculated with the reduced theory 
(\ref{eq:13}) with the same quantity obtained in the high temperature 
4-dimensional Yang-Mills theory yields values for the effective 
coupling constants $\beta $, $h$ and $\kappa $. The values which we will 
use below were taken from refs.~\cite{lac92,kar94} (see table \ref{tab:1}). 

\vskip 0.3cm 
Interpreting the effective field theory acting in the spatial hypercube 
of the full 4-dimensional theory as a field theory in $2+1$ dimensions, 
one considers the potential calculated from spatial Wilson loops as the 
{\it spatial} static quark anti-quark potential. 
The derivative of these potential yields the spatial string tension for 
asymptotic values of the distance $r$. Using the effective theory 
(\ref{eq:13}) and the parameters of table (\ref{tab:1}), we calculated 
the derivative of the spatial static potential. We compare these data 
with the result for the spatial static potential calculated with center 
projected configurations (see previous section) in figure \ref{fig:3}. 
An agreement of the spatial string tension calculated from the full 
and the projected configurations, respectively, as well as a consistency 
with the asymptotic value (\ref{eq:2}) of the 4-dimensional theory 
is observed. Comparing figure \ref{fig:3} with figure \ref{fig:1}, 
we conclude that the adjoint Higgs field yields minor corrections 
to the spatial static potential. 

\vskip 0.3cm 
\begin{figure}[t]
\centerline{ 
\epsfxsize=10cm
\epsffile{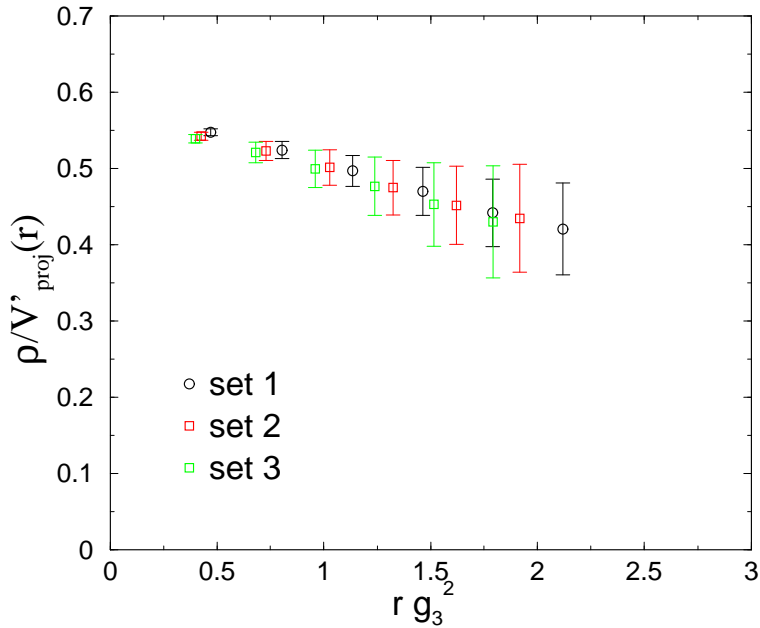}
}
\caption{ The ratio of the  vortex area density $\rho $ and the 
   derivative of the projected quark anti-quark potential at several 
   temperatures (see table \ref{tab:1}). } 
\label{fig:4} 
\end{figure}
Finally, we have calculated the ratio of vortex (area) density 
and (spatial) string tension at temperatures provided by the 
parameter sets of table \ref{tab:1} using the effective theory 
(\ref{eq:13}). The result is shown in figure \ref{fig:4}. 
Comparing the scale of the horizontal axis in figure \ref{fig:2} and figure 
\ref{fig:4}, we conclude that the above ratio has not yet reached 
its asymptotic value for the parameters given in table~\ref{tab:1}.

\vskip 0.3cm 
\centerline{ \bf 5. Conclusions: \hfill } 

We have studied the spatial string tension in 3-dimen\-sional {\it pure} 
Yang-Mills theory and in the 3-dimensional Yang-Mills theory 
coupled with adjoint Higgs matter. The latter model applies 
as the dimensional reduced theory describing the high temperature 
phase of 4-dimensional $SU(2)$ Yang-Mills theory~\cite{lac92,kar94}. 

\vskip 0.3cm 
In both cases, we compare the (spatial) string tension of the full 
simulation with the value obtained when the link configurations 
are reduced to vortex configurations by center projection. 
We find that the (spatial) string tension is approximated by the vortex 
configurations within a numerical accuracy of 10\%. The numerical error 
mainly results from the gauge fixing procedure and is largely due 
to the average over Gribov copies. A modification of the gauge 
fixing condition~\cite{ale99,la00} is advisable for improving 
the numerical accuracy.

\vskip 0.3cm 
Furthermore, our results indicate that the vortex area density of 
3-dimensional pure $SU(2)$ gauge theory extrapolates to the continuum limit 
of vanishing lattice spacing. The vortex area density is only slightly 
changed by the coupling of the 3-dimensional Yang-Mills theory to 
the adjoint Higgs matter. Using these results, we 
estimate that $\rho / \sigma _s \approx 0.38 \pm 0.12 $ which is consistent
with the estimate $0.33 \pm 0.05$ obtained from a calculation 
using the full 4-dimensional theory. 

\vskip 0.3cm 
Our results support the vortex picture of the high 
temperature phase of 4-dimensional 
Yang-Mills theory: the center vortices of the latter theory are aligned 
along the time axis direction by temperature effects~\cite{la99}, while their 
fingerprint in the spatial hypercube constitutes a percolating 
vortex cluster. Hence, we find first evidence that the findings of dimensional 
reduction~\cite{app81,lac92,kar94,bali93} extend to the vortex picture.

\vskip 1cm

{\bf Note added: }

During the preparation of this manuscript, ref.~\cite{har00} 
appeared. In that work, the 't~Hooft twisted 
vortices were investigated in 3-dimensional $SU(2)$ Yang-Mills theory. 
These investigations are complementary to the studies of the present 
letter, in which the dynamical properties of the center vortices 
which emerge in center gauge 
by center projection have been studied.

\end{document}